\pdfoutput=1
\RequirePackage{ifpdf}
\ifpdf % We~are running pdfTeX in pdf mode
\documentclass[pdftex]{sigma}
\else
\documentclass{sigma}
\fi

\usepackage{mathrsfs}

\numberwithin{equation}{section}

\begin{document}

\allowdisplaybreaks

\newcommand{\arXivNumber}{2508.04158}

\renewcommand{\PaperNumber}{005}

\FirstPageHeading

\ShortArticleName{On Integrable Structure of Null String in (Anti-)de Sitter Space}

\ArticleName{On Integrable Structure of Null String\\ in (Anti-)de Sitter Space}

\Author{Dmytro V.~UVAROV}

\AuthorNameForHeading{D.V.~Uvarov}

\Address{NSC Kharkiv Institute of Physics and Technology, 1~Academichna Str., Kharkiv, Ukraine}
\Email{\mail{d\_uvarov@hotmail.com}}

\ArticleDates{Received August 07, 2025, in final form January 06, 2026; Published online January 16, 2026}

\Abstract{Presently integrability turned out to be the key property in the study of duality between superconformal gauge theories and strings in anti-de Sitter superspaces. Complexity of the study of integrable structure in string theory is caused by complicated dependence of background fields of the Type~II supergravity multiplets, with which strings interact, on the superspace coordinates. This explains an interest to study of limiting cases, in which superstring equations simplify. In the present work, we considered the limiting case of zero tension corresponding to null string. The representation in the form of the Lax equation of null-string equations in (anti-)de Sitter space realized as a coset manifold is obtained. Proposed is twistor interpretation of the Lagrangian of (null) string in anti-de Sitter space expressed in terms of group variables.}


\Classification{81T30; 81T35; 83C80; 37K10; 83C60}

\section{Introduction}

Construction of consistent quantum theory that would unify gravity with the Standard model remains an unsolved problem. String theory, which basic constituents are extended relativistic objects such as strings and branes, offers an interesting direction for seeking a solution of this problem. Application of the holographic principle in string theory led to hypothesis of the AdS/CFT correspondence that suggests formulation of the quantum gravity in anti-de Sitter superspaces on the basis of dual gauge theories on their conformal boundary
\cite{Gubser1998,Maldacena1997, Witten1998}.

However, it is also well known that the problem of finding energy spectrum even of free superstrings in curved superspaces appears extremely difficult. It is explained by complicated highly non-linear structure of dynamical equations of superstrings caused by interplay of the background geometry and elastic force of the world sheet proportional to string tension. Possible simplified approach consists in quantization of small oscillations around particular classical solutions of superstring equations (see, e.g., review \cite{Tseytlin2010}). Another approach is to consider models of null (super)strings that are extended objects with zero tension \cite{Karlhede1986,Schild1977,Zheltukhin1987,Zheltukhin1988a,Zheltukhin1988b}. In curved backgrounds, their dynamics is less complicated because of the absence of elastic force~\cite{Blair2023,Bonelli2003,Dabrowski1996,Dabrowski2002,deVega1992,Fursaev2017,Kar1995,Porfyriadis1997,Roshchupkin1995}. It was suggested to consider null strings as the leading order approximation in the perturbative approach to solution of the string equations in curved spaces~\cite{deVega1992,Zheltukhin1996}.

Apart from perturbative approaches to solve string equations, applicable to a wide class of curved spaces \cite{deVega1992,Roshchupkin1998,Zheltukhin1996,Roshchupkin1997}, for certain backgrounds with high symmetry, in particular cosets of (semisimple) Lie groups, there was proved classical integrability of respective two-dimensional sigma-models. Their equations were presented in the form of the zero-curvature condition for the Lax connection 1-form that depends on the spectral parameter \cite{Eichenherr1979,Luscher1978,Pohlmeyer1976}. Integrability of two-dimensional sigma-models and strings on various group manifolds was examined also in~\cite{Barbashov1981,Zakharov1978,Zheltukhin1983}. It implies existence of an infinite number of the world-sheet conserved \mbox{(non-)local} currents. There were elaborated powerful methods to find the spectrum of such models (see, e.g., reviews \cite{Bombardelli2016,Zarembo2017}). Especially interesting among these coset spaces are de Sitter and anti-de Sitter spaces, that are solutions of the vacuum Einstein equations with positive/negative cosmological constant. They can be realized as the coset manifolds
\[
\mathrm{SO}(1,D)/\mathrm{SO}(1,D-1)
\qquad \text{and}\qquad {\mathrm{SO}(2,D-1)/\mathrm{SO}(1,D-1)},\] respectively. Integrable structures of two-dimensional sigma-models in these spaces have been extensively examined (see, e.g., \cite{Barbashov1981,Burrington2009,deVega1993,Hoare2012,Katsinis2022}).

Presently, interest to supersymmetric integrable two-dimensional sigma-models and strings, in particular in anti-de Sitter superspaces is mainly connected with the above mentioned application of the holographic principle in string theory. It was triggered by proof of the classical integrability~\cite{Bena2003} of equations of two-dimensional sigma-model on the $\mathrm{PSU}(2,2|4)/(\mathrm{SO}(1,4)\times \mathrm{SO}(5))$ supercoset manifold. The latter is used to describe dynamics of the IIB superstring in the~${\mathrm{AdS}_5\times S^5}$ superspace \cite{Kallosh1998,Metsaev1998}. Integrable structure of the superstring correlates with that of dual $D=4$ $\mathcal N=4$ supersymmetric Yang--Mills theory, initially studied in \cite{Minahan2002}. Moreover, this motivated the search of integrable structures also in the lower-dimensional examples of the AdS/CFT correspondence (see, e.g., reviews \cite{Demulder2023,Klose2010}). However, on the string side of dualities their identification and study are complicated by insufficiently high symmetries of relevant 10-dimensional anti-de Sitter superspaces as opposed to the $\mathrm{AdS}_5\times S^5$ superspace that is maximally supersymmetric solution of the $D=10$ chiral supergravity constraints.

This problem manifests itself already in the case of the $\mathrm{AdS}_4/\mathrm{CFT}_3$ correspondence \cite{Aharony2008}. This is the instance of duality between gauge fields and strings, in which the IIA superstring theory in the $\mathrm{AdS}_4\times\mathbb{CP}^3$ superspace, that breaks 8 of 32 supersymmetries, \cite{Gomis2008} is described as the superconformal Chern--Simons-matter gauge theory in three dimensions. In \cite{Arutyunov2008,Stefanski2008}, there has been constructed two-dimensional $\mathrm{OSp}(4|6)/(\mathrm{SO}(1,3)\times \mathrm{U}(3))$ supercoset sigma-model along the same lines as the $\mathrm{PSU}(2,2|4)/(\mathrm{SO}(1,4)\times \mathrm{SO}(5))$ one. However, it correctly describes only a~subsector of the $\mathrm{AdS}_4\times\mathbb{CP}^3$ superstring dynamics \cite{Arutyunov2008,Gomis2008,Stefanski2008}. In the domain of applicability of the $\mathrm{OSp}(4|6)/(\mathrm{SO}(1,3)\times \mathrm{U}(3))$ sigma-model, the classical integrability of its equations was proved in \cite{Arutyunov2008,Stefanski2008} by extending the argument of \cite{Bena2003}. To prove integrability of the complete set of dynamical equations of the $\mathrm{AdS}_4\times\mathbb{CP}^3$ superstring, one has to find extension of the Lax connection of the $\mathrm{OSp}(4|6)/(\mathrm{SO}(1,3)\times \mathrm{U}(3))$ sigma-model by contributions of eight Grassmann coordinates for broken supersymmetries of the $\mathrm{AdS}_4\times\mathbb{CP}^3$ superspace. In the absence of the systematic approach for proving classical integrability of dynamical equations, attempts were made to find such an extension order by order in these eight coordinates \cite{Cagnazzo2011,Sorokin2010}, in particular using the $\kappa$-symmetry gauge freedom \cite{Uvarov2012a}.

This explains an interest to study the limiting cases, in which the $\mathrm{AdS}_4\times\mathbb{CP}^3$ superstring equations are simplified, and proof of their integrability becomes feasible. One of such limits corresponds to infinite tension of the superstring, in which it shrinks to a point superparticle. Integrability of its equations in the $\mathrm{AdS}_4\times\mathbb{CP}^3$ superspace has been proved in \cite{Uvarov2012b,Uvarov2012d}. Moreover, it has been shown that the Lax pair that enters the Lax representation for the superparticle's equations retains some information about structure of the Lax connection of the tensile superstring.

Examined here is the opposite limit of zero tension that corresponds to the null string, and the Lax representation of its equations is found. We focus on the world-sheet structure of this Lax representation. So, we restrict ourselves to the case of bosonic null string in the (anti-)de Sitter space. Integrability of its equations is anticipated in view of integrability of tensile string equations \cite{Barbashov1981,Zheltukhin1983} and the fact that null string equations in conformally-flat space-times are exactly solvable \cite{Roshchupkin1995,Roshchupkin1998}.

\section{Formulations of null strings in flat and curved spaces}

Null string Lagrangian was proposed by A.~Schild \cite{Schild1977}
$
\mathscr L(\xi)=\frac12 g(\xi)$,
where $g=\det g_{ij}$ is the determinant of induced world-sheet metric $g_{ij}(\xi)=\partial_i X^{m'}(\xi)g_{m'n'}(X(\xi))\partial_j X^{n'}(\xi)$, $g_{m'n'}(X)$ is the metric of (curved) space-time with coordinates $X^{m'}$ and $\xi^i=(\tau,\sigma)$ are local world-sheet coordinates. It
contains the first power of the determinant of induced world-sheet metric rather than its square root as in the case of the Nambu--Goto string. Schild's Lagrangian was generalized in \cite{Karlhede1986,Zheltukhin1987,Zheltukhin1988a,Zheltukhin1988b} by introducing world-sheet scalar density
$E(\xi)$ of weight $-1$ that multiplies this determinant
\begin{gather}\label{-1}
\mathscr L(\xi)=\frac{g(\xi)}{2E(\xi)}
\end{gather}
rendering the null string action
$
S=\int d^2\xi\mathscr L(\xi)
$
reparametrization invariant. In such a formulation, Lagrangian equations for the space-time coordinate fields are non-linear.

Another formulation of the tensionless string \cite{Lindstrom1991}
\begin{gather}\label{0}
S=\int d^2\xi\mathscr L(\xi),\qquad\mathscr L(\xi)=
-
\frac12\rho^i\rho^j g_{ij}
\end{gather}
includes a pair of auxiliary fields that make up world-sheet vector density $\rho^i(\xi)$.
Like in the Polyakov formulation \cite{Brink1976,Deser1976,Polyakov1981}
\begin{gather}\label{0'}
S=\int d^2\xi\mathscr L(\xi),\qquad\mathscr L(\xi)=-\frac{T}{2}\sqrt{-\gamma}\gamma^{ij}g_{ij},
\end{gather}
equations for the space-time coordinate fields are linear.

Other formulations of the null string (and brane) Lagrangian that additionally include tangent to the world-sheet components of the local Cartan frame were examined in \mbox{\cite{Bandos1990,Bandos1991a,Bandos1993a,Zheltukhin1987,Zheltukhin1988a,Zheltukhin1988b}}.
They are classically equivalent to those mentioned above and were generalized to null superstrings (and superbranes) resulting in irreducible realization of the $\kappa$-symmetry of the action for arbitrary amount of the space-time supersymmetry.

All these formulations of the null string arise as the tension-to-zero limit of respective tensile string formulations. The Lagrangian \eqref{-1} is the limiting case of the Nambu--Goto string reformulation \cite{Zheltukhin1987,Zheltukhin1988a,Zheltukhin1988b}
\begin{gather}\label{-1'}
S=\int d^2\xi\mathscr L(\xi),\qquad\mathscr L(\xi)=\frac{g(\xi)}{2E(\xi)}-\frac{T^2}{2}E(\xi)
\end{gather}
that allows us to take the tension-to-zero limit in the same way as one takes massless limit in the massive point particle model
\begin{gather}\label{-1''}
S=\int d\tau\mathscr L(\tau),\qquad\mathscr L(\tau)=\frac{\dot X^2}{2e(\tau)}-\frac{m^2}{2}e(\tau)\;\xrightarrow[m\to0]{}\;\mathscr L(\tau)=\frac{\dot X^2}{2e(\tau)}.
\end{gather}

The formulation \eqref{0} can be derived either from the Nambu--Goto or the Polyakov Lagrangians. To this end one expresses the Lagrangian in terms of the space-time coordinates and momenta in order to get parametrization of auxiliary world-sheet metric appropriate for the limiting transition, then integrates out the momenta and takes the limit. Also the limit can be taken in the phase-space formulation resulting in the null string Lagrangian expressed via canonical variables \cite{Gamboa1989,Gamboa1990,Zheltukhin1987,Zheltukhin1988a,Zheltukhin1988b}, from which \eqref{0} can be obtained. The details can be found in the original work \cite{Lindstrom1991} and have been recapitulated more recently in \cite{Casali2016}. Both Polyakov formulation \eqref{0'} and that with auxiliary scalar density \eqref{-1'} can be derived from the unified formulation of \cite{Kato1983}.

Analogously null (super)string formulations including local Cartan frame components \cite{Bandos1990,Bandos1991a,Bandos1993a,Zheltukhin1987,Zheltukhin1988a,Zheltukhin1988b} can be derived from respective tensile (super)string formulations \cite{Bandos1991b,Bandos1994,Volkov1985} (see, e.g., \cite{Bandos2014}).

The opposite limit of infinite tension is also of considerable interest since it produces massless (super)particle models. Their dynamics is much simpler than that of (super)strings and quantum spectrum contains finite number of the lowest (super)string states. When string tension becomes very strong, its oscillations are damped that can be expressed as $\partial X^{m'}(\tau,\sigma)/\partial\sigma=0$. This condition leaves only zero modes in the Fourier expansion of the space-time coordinate fields in~$\sigma$. It is used in reduction procedure of a $(p+1)$-brane to $p$-brane (see, e.g., \cite{Duff1987}). Then there remains just one term in the Polyakov Lagrangian \eqref{0'}
\begin{displaymath}
\mathscr L(\xi)\vert_{\partial X^{m'}(\tau,\sigma)/\partial\sigma=0}=-\frac{T}{2}\sqrt{-\gamma}\gamma^{\tau\tau}\dot X^{2}.
\end{displaymath}
So that integrating in $\sigma$ and taking the limit $T\to\infty$ gives massless particle action \eqref{-1''}, in which the Lagrange multiplier $e(\tau)=-\lim_{T\to\infty}Tl_s\sqrt{-\gamma}\gamma^{\tau\tau}$ and the string length $l_s\sim T^{-1/2}$.
Note that there can be defined different procedures to take the infinite tension limit that produce from the Polyakov string not only massless particle but also the ambitwistor string \cite{Berkovits2013,Mason2013} and even the null string \cite{Bandos2014}.

In the above discussion, the background on which (null) string propagates has not been specified. Below we specialize to the case of coset manifolds and, in particular, of the (anti-)de Sitter space. There vielbein 1-form is identified with the Cartan forms associated with generators of the quotient algebra. As a result, Lagrangian equations of the null string in these spaces in the formulation \cite{Lindstrom1991} take the form of the first-order partial differential equations for the Cartan forms similarly to the case of two-dimensional sigma-models \cite{Eichenherr1979,Zheltukhin1981,Zheltukhin1982} and tensile superstrings~\cite{Bandos1993b,Bena2003}. In the next section, we present them as the first-order differential Lax equation for the Lax pair that takes value in the symmetry algebra of the group manifold similarly to the Lax connection of two-dimensional sigma-models \cite{Eichenherr1979,Luscher1978,Pohlmeyer1976,Zakharov1978} and tensile strings~\cite{Barbashov1981,Zheltukhin1983}.
Since the Hamiltonian formalism plays important role in study of integrable models and original derivation of the null string Lagrangian of \cite{Lindstrom1991} uses it,
we also consider the first-order representation for the null string Lagrangian that includes components of the momentum density conjugate to the space-time coordinates. We obtain equations of the null string and expressions for the Lax pair components in terms of these phase-space variables. Then the Lax representations in configurational space and phase space are compared with those for tensile string in the (anti-)de Sitter space-time.

\section[Lax representation of null string equations in (anti-)de Sitter space]{Lax representation of null string equations\\ in (anti-)de Sitter space}

\subsection{Group-theoretic description of geometry of (anti-)de Sitter space}

Relations of the $\mathfrak{so}(2,D-1)$ isometry algebra of the $D$-dimensional anti-de Sitter space and of the $\mathfrak{so}(1,D)$ isometry algebra of the $D$-dimensional de Sitter space can be presented in the uniform way
$
[M_{ab},M_{cd}]=\eta_{ad}M_{bc}-\eta_{ac}M_{bd}
-\eta_{bd}M_{ac}+\eta_{bc}M_{ad}$,
where $-\eta_{00}=\eta_{11}=\dots=\eta_{D-1D-1}=1$ and $\eta_{DD}=s$ with $s=-1$ for anti-de Sitter space and $+1$ for de Sitter space. These relations in the form, in which generators of the $\mathfrak{so}(2,D-1)/\mathfrak{so}(1,D-1)$ ($\mathfrak{so}(1,D)/\mathfrak{so}(1,D-1)$) quotient algebra are manifestly separated, are known as~$(\mathfrak{a})\mathfrak{ds}_D$ algebra
\begin{gather*}
[M_{Db'},M_{Dd'}]=-sM_{b'd'},\qquad
[M_{a'b'},M_{Dd'}]=-\eta_{a'd'}M_{Db'}+\eta_{b'd'}M_{Da'}, \\
[M_{a'b'},M_{c'd'}]=\eta_{a'd'}M_{b'c'}-\eta_{a'c'}M_{b'd'}-\eta_{b'd'}M_{a'c'}+\eta_{b'c'}M_{a'd'},\\ a',b',c',d'=0,\dots, D-1.
\end{gather*}
Define Cartan 1-form with value in the $(\mathfrak{a})\mathfrak{ds}_D$ algebra
\begin{gather}\label{1}
\mathcal C(d)=\mathscr G^{-1}d\mathscr G=2G^{Da'}(d)M_{Da'}+G^{a'b'}(d)M_{a'b'}\in (\mathfrak{a})\mathfrak{ds}_D,
\end{gather}
where $\mathscr G\in \mathrm{SO}(2,D-1)/\mathrm{SO}(1,D-1)$ for anti-de Sitter space and $\mathscr G\in \mathrm{SO}(1,D)/\mathrm{SO}(1,D-1)$ for de Sitter space. By definition, the Cartan form obeys the Maurer--Cartan equation that in terms of differential forms is written as $d\mathcal C(d)=\mathcal C(d)\wedge\mathcal C(d)$. It splits into equations for the Cartan forms associated with generators of the $\mathfrak{so}(2,D-1)/\mathfrak{so}(1,D-1)$ ($\mathfrak{so}(1,D)/\mathfrak{so}(1,D-1)$) quotient algebra
\begin{gather}\label{2}
dG^{Da'}(d)=2G^{a'}{}_{b'}(d)\wedge G^{Db'}(d)
\end{gather}
and equations for the Cartan forms associated with the $\mathfrak{so}(1,D-1)$ stability algebra generators
\begin{gather}\label{3}
dG^{a'b'}(d)=-sG^{Da'}(d)\wedge G^{Db'}(d)+sG^{Db'}(d)\wedge G^{Da'}(d)+2G^{a'}{}_{c'}(d)\wedge G^{c'b'}(d).
\end{gather}
The Cartan forms $G^{Da'}(d)$ are identified with the vielbein 1-form $E^{a'}(d)=dX^{m'}E^{a'}_{m'}(X)$ of the (anti-)de Sitter space parametrized by (local) coordinates $X^{m'}$ and $G^{a'b'}(d)$ is identified with its spin connection 1-form $\Omega^{a'b'}(d)=dX^{m'}\Omega^{a'b'}_{m'}(X)$ as
$E^{a'}(d)=G^{Da'}(d)$, $\Omega^{a'b'}(d)=2G^{a'b'}(d)$.
The coefficient +2 in the second equality follows from comparison of the Maurer--Cartan equation~\eqref{2} with the zero-torsion condition for the space-time vielbein
\[
T^{a'}(d)=dE^{a'}(d)-\Omega^{a'}{}_{b'}(d)\wedge E^{b'}(d)=0.
\]
Identification of the Cartan forms with the vielbein and spin connection transforms the Maurer--Cartan equation~\eqref{3} into the definition of the curvature 2-form of the (anti-)de Sitter space
\[
R^{a'b'}(d)=d\Omega^{a'b'}(d)-\Omega^{a'}{}_{c'}(d)\wedge\Omega^{c'b'}(d)=-2sE^{a'}(d)\wedge E^{b'}(d).
\]

\subsection{Lagrangian formulation of null string in (anti-)de Sitter space}

The action of the null string in (anti-)de Sitter space
\begin{gather}\label{4}
S=\int d^2\xi\mathscr L(\xi),\qquad\mathscr L(\xi)=
-
\frac12\rho^i\rho^j G^{Da'}_i\eta_{a'b'}G^{Db'}_j,
\end{gather}
depends on the world-sheet projections of the Cartan forms associated with the quotient algebra generators $G^{Da'}_i=\partial_i X^{m'}(\xi)E^{a'}_{m'}(X(\xi))$. Induced world-sheet metric in terms of Cartan forms is defined as
\[
g_{ij}(\xi)=G^{Da'}_i\eta_{a'b'}G^{Db'}_j=\partial_i X^{m'}g_{m'n'}(X)\partial_j X^{n'},
\]
 where $g_{m'n'}(X)=E^{a'}_{m'}(X)\eta_{a'b'}E^{b'}_{n'}(X)$ is the metric tensor of (anti-)de Sitter space and $\eta_{a'b'}$ is the tangent space Minkowski metric proportional to the Killing form of the $\mathfrak{so}(2,D-1)$ ($\mathfrak{so}(1,D)$) algebra restricted to the subspace spanned by generators of the quotient algebra. Due to the presence of $\eta_{a'b'}$, the space-time and induced metrics are invariant under local $\mathrm{SO}(1,D-1)$ symmetry generated by stability group action on the coset representative $\mathscr G'=\mathscr GH$, $H\in \mathrm{SO}(1,D-1)$. They are also invariant under global $\mathrm{SO}(2,D-1)$ ($\mathrm{SO}(1,D)$) symmetry $\mathscr G'H^{-1}=G\mathscr G$, $ G\in \mathrm{SO}(2,d-1)$ ($\mathrm{SO}(1,D)$) of (anti-)de Sitter space, but the Cartan forms get transformed as follows from their definition~\eqref{1}%
\begin{gather}\label{5}
\mathcal C'(d)=H^{-1}\mathcal C(d)H+H^{-1}dH.
\end{gather}
In particular, $G^{a'b'}(d)$ transforms as 1-form connection that justifies its identification with the spin connection of (anti-)de Sitter space-time.

To make action \eqref{4} reparametrization invariant, the auxiliary world-sheet vector densi\-ty~$\rho^{i}(\xi)$ should have weight $w=-1/2$. (This vector density is proportional to the world-sheet zweibein components and enters formulations of both tensionless and tensile (super)strings proposed and examined in \cite{Bandos1990,Bandos1991a,Bandos1991b,Bandos1994}.) So under world-sheet reparametrizations $\xi'=\xi'(\xi)$ it transforms as
\[
\rho'^{i}(\xi')=J^{-w}\frac{\partial\xi'^{i}}{\partial\xi^{j}}\rho^j(\xi), \qquad J=\left\vert{\det \frac{\partial\xi}{\partial\xi'}}\right\vert.
\]

Algebraic equations resulting from variation of $\rho^i(\xi)$
$
g_{ij}\rho^j=0
$
imply that $\rho^i$ is non-zero eigenvector of the induced metric $g_{ij}$ with zero eigenvalue that amounts to vanishing of its determinant.

\subsection[Dynamical equations of null string in (anti-)de Sitter space and their Lax representation]{Dynamical equations of null string in (anti-)de Sitter space\\ and their Lax representation}

Equations for the space-time coordinate fields $X^{m'}(\xi)$ can be presented through the Cartan forms if one takes not $\delta X^{m'}(\xi)$, but $G^{Da'}(\delta)=\delta X^{m'}E^{a'}_{m'}$ as variation parameters. The latter can be obtained from the Cartan forms $G^{Da'}(d)$ by applying the operator of formal substitution~${i_\delta G^{Da'}(d)=\delta X^{m'}E^{a'}_{m'}}$. One also has to use well-known Cartan formula for the Lie derivative of a differential form $F(d)$: $\mathcal LF(d)=di_\delta F+i_\delta dF$. It equals variation of those differential forms, which dependence on the space-time coordinates and their differentials does not change. This holds, in particular, for background fields. When applied to the Cartan forms $G^{Da'}(d)$, one should substitute in the second summand the Maurer--Cartan equations \eqref{2}. This results in the following equation:
\begin{gather}\label{6}
%-
\frac{\delta S}{\delta G^D{}_{a'}(\delta)}=\rho^i\nabla_i\bigl(\rho^j G^{Da'}_j\bigr)+2\rho^i G_i^{a'}{}_{b'}\rho^j G^{Db'}_j=0,
\end{gather}
where $\nabla_i$ is a covariant derivative on the world-sheet and vector density $\rho^i$ is taken to be divergence-free $\nabla_i\rho^i=0$. Since $\rho^i$ plays the role of zweibein on the null world sheet, this condition is an analogue of the covariant constancy condition for metric/zweibein on the non-singular world-sheet \cite{Lindstrom1991}.

We write Lagrangian equation \eqref{6} as the Lax equation
\begin{gather}\label{7}
\rho^i\nabla_i L-[L,M]=0,
\end{gather}
where
\smash{$
L=\rho^i G^{Da'}_i M_{Da'}$}, \smash{$ M=\rho^i G^{a'b'}_i M_{a'b'}$}
are components of the Lax pair. Note that the Lax equation is invariant under local $\mathrm{SO}(1,D-1)$ stability group transformation that acts on the Lax pair components in the following way
$
L'=H^{-1}LH$, $
M'=H^{-1}MH+H^{-1}\rho^i\nabla_i H
$
in accordance with the transformation law \eqref{5} for the Cartan forms. The Lax representation~\eqref{7} has the same form as in the case of massless (super)particle \cite{Uvarov2012b,Uvarov2012d} modulo substitution $\rho^i\nabla_i\rightarrow\frac{d}{d\tau}$, where $\tau$ is the (super)particle's world-line parameter.
This implies that the null string effectively behaves like one-dimensional dynamical system though the null world sheet is two-dimensional. As is known, for one-dimensional integrable systems the Lax representation may not contain spectral parameter. For two-dimensional integrable systems, in particular, tensile (super)strings spectral parameter is responsible for generation of the infinite number of (non-)local conserved currents. Therefore its absence in the Lax equation \eqref{7} may indicate that in the limit of zero tension (two-dimensional metric degeneration) there remains finite number of conserved local currents. At the same time definition of the zero-tension limiting transition on the level of the Lax representation of the string equations and the integrable structure requires further study both on the classical and quantum level.

\subsection[Lax representation of Hamiltonian equations of null string in (anti-)de Sitter space]{Lax representation of Hamiltonian equations of null string\\ in (anti-)de Sitter space}

In the original derivation of the null string Lagrangian in the formulation with auxiliary vector density \eqref{4},
transition to the phase-space formulation of the tensile string is part of the procedure to take the tension-to-zero limit. It is also known that the Hamiltonian formalism plays important role in the study of integrable systems. This motivates us to work out the Lax representation for dynamical equations that follow from the null string Lagrangian expressed in terms of the phase-space variables.

To this end, introduce density of the canonical momentum conjugate to the space-time coordinates
\begin{gather}\label{8}
p_{m'}(\tau,\sigma)=-\rho^\tau\rho^\tau g_{m'n'}(X)\partial_\tau X^{n'}-\rho^\tau\rho^\sigma g_{m'n'}(X)\partial_\sigma X^{n'}.
\end{gather}
It is convenient to introduce momentum density with the tangent-space index
$
 p_{a'}(\tau,\sigma)=E^{n'}_{a'}p_{n'}$,
where $E^{n'}_{a'}$ is the inverse space-time vielbein: $E^{m'}_{a'}E_{m'}^{b'}=\delta_{a'}^{b'}$. Then Lagrangian of the null string is written in the form
\begin{gather}
\mathscr L(\tau,\sigma)=p_{m'}\partial_\tau X^{m'}-\mathscr H=p_{a'}G^{Da'}_\tau-\mathscr H,\nonumber\\
-
\mathscr H(\tau,\sigma)=\frac{1}{2(\rho^\tau)^2}p_{a'}\eta^{a'b'}p_{b'}+\frac{\rho^\sigma}{\rho^\tau}p_{a'}G^{Da'}_\sigma,\qquad\rho^\tau\not=0. \label{9}
\end{gather}
As should be for dynamical systems invariant under reparametrization symmetry, density of the Hamiltonian is given by the sum of the first-class constraints
\begin{gather}\label{10}
p^2=p_{a'}\eta^{a'b'}p_{b'}\approx0,\qquad p_{a'}G^{Da'}_\sigma\approx0,
\end{gather}
that are generators of this symmetry, with arbitrary Lagrange multipliers. Their role is played by components of the world-sheet vector density $\rho^i$ like in the case of tensile string \cite{Bandos1994}. Variation of the Lagrangian \eqref{9} on the phase-space variables gives two equations
\begin{gather}
\rho^i G^{Da'}_i+\frac{1}{\rho^\tau}p^{a'}=0,\label{11}
\\
\rho^i\nabla_i\left(\frac{1}{\rho^\tau}p^{a'}\right)+2\rho^i G_i^{a'}{}_{b'}\frac{1}{\rho^\tau}p^{b'}=0.\label{12}
\end{gather}
Equation~\eqref{11} determines the momentum density and equation~\eqref{12} is dynamical equation. When writing these equations, it was taken into account that $(\rho^\tau)^{-1}p^{a'}$ transforms under reparametrizations as a world-sheet scalar density of weight $w=-1/2$ as follows from the definition of momentum density \eqref{8}
\begin{displaymath}
\frac{1}{\rho^\tau}p_{m'}(\tau,\sigma)=-\rho^i g_{m'n'}(X)\partial_i X^{n'}.
\end{displaymath}

As is known, covariant derivative of a tensor density $\phi$ of weight $w$ includes contribution of contracted Christoffel symbols
$
\nabla_i\phi^{\cdots}_{\cdots}=\partial_i\phi^{\cdots}_{\cdots}+w\Gamma^j_{ij}\phi^{\cdots}_{\cdots}+\cdots$,
where dots stand for terms that are the same for tensor density and respective true tensor. In the case of the null string, the condition $\nabla_i\rho^i=0$ determines projection of $\Gamma^j_{ij}$ on $\rho^i$: $\rho^i\Gamma^j_{ij}=-2\partial_i\rho^i$. This relation allows us to write equation~\eqref{12} in another form
\begin{displaymath}
\partial_i\left(\frac{\rho^i}{\rho^\tau}p^{a'}\right)+2\rho^i G_i^{a'}{}_{b'}\frac{1}{\rho^\tau}p^{b'}=0.
\end{displaymath}
Equation~\eqref{12} can be cast into the form of the Lax equation \eqref{7}, in which in the expression for the Lax component $L$ equation~\eqref{11} has been substituted
\begin{gather}\label{12'}
L=\rho^i G^{Da'}_i M_{Da'}=-\frac{1}{\rho^\tau}p^{a'}M_{Da'}.
\end{gather}

\subsection[Lax representation of Lagrangian and Hamiltonian equations of tensile string in (anti-)de Sitter space]{Lax representation of Lagrangian and Hamiltonian equations\\ of tensile string in (anti-)de Sitter space}

While studying tension-to-zero limit of the integrable structure of string in (anti-)de Sitter space, it is of interest to compare the obtained Lax representation for the null string equations with that in the case of non-zero tension.
First we recapitulate the zero-curvature representation of the Lagrangian equations of tensile string. Then consider the Hamiltonian equations because their realization as the zero-curvature condition is suitable for comparison with the Lax equation for the null string.

Tensile string counterpart of the null string action \eqref{4} is well known
\begin{gather}\label{13}
S=\int d^2\xi\mathscr L(\xi),\qquad\mathscr L(\xi)=-\frac{T}{2}\sqrt{-\gamma}\gamma^{ij} G^{Da'}_i\eta_{a'b'}G^{Db'}_j.
\end{gather}
It includes non-singular auxiliary world-sheet metric $\gamma_{ij}$ with determinant $\gamma$ and inverse metric~$\gamma^{ij}$. It is also known representation of the string Lagrangian in terms of differential forms
\begin{gather}\label{14}
\mathscr L(\xi)=\frac{T}{2}G^{Da'}(d)\eta_{a'b'}\wedge\ast G^{Db'}(d),
\end{gather}
where the Hodge dual of a world-sheet 1-form $\alpha(d)=d\xi^i\alpha_i$ is defined as
\begin{gather}
\ast\alpha(d)=-\sqrt{-\gamma}d\xi^k\varepsilon_{kj}\gamma^{ji}\alpha_i=-\frac{1}{\sqrt{-\gamma}}d\xi^k\gamma_{kj}\varepsilon^{ji}\alpha_i,\\
\varepsilon^{ij}=-\varepsilon^{ji},\qquad \varepsilon^{ij}\varepsilon_{jk}=\delta^i_k,\qquad \varepsilon^{\tau\sigma}=+1.
\end{gather}
From action \eqref{13} with the Lagrangian expressed in terms of differential forms \eqref{14}, there follow dynamical equations of the tensile string
\begin{displaymath}
d\ast G^{Da'}(d)-2G^{a'}{}_{b'}(d)\wedge\ast G^{Db'}(d)=0,
\end{displaymath}
which form is similar to the Maurer--Cartan equation \eqref{2}. This similarly underlies the Lax representation of these equations as the zero curvature condition
\begin{gather}\label{15}
dL(d)-L(d)\wedge L(d)=0
\end{gather}
of the Lax 1-form
\begin{displaymath}
 L(d)=2L^{Da'}(d)M_{Da'}+L^{a'b'}M_{a'b'}\in(\mathfrak{a})\mathfrak{ds}_D,
\end{displaymath}
where
\smash{$
L^{Da'}(d)=\ell_1 G^{Da'}(d)+\ell_2\ast G^{Da'}(d)$}, \smash{$ L^{a'b'}(d)=G^{a'b'}(d)$}.
The dependence of functions $\ell_1$ and $\ell_2$ on the spectral parameter is determined by algebraic equation $\ell_1^2-\ell_2^2=1$ that is the consequence of the zero curvature condition. Equation~\eqref{15} is invariant under gauge transformation of the Lax 1-form $L'(d)=H^{-1}L(d)H+H^{-1}dH$, $H\in \mathrm{SO}(1,D-1)$ that allows us to interpret it as trivial $\mathrm{SO}(1,D-1)$ connection. Let us also note that in the case, when the world-sheet reparametrizations and dilations are used to bring auxiliary metric equal to the two-dimensional Minkowski metric $\gamma_{ij}=\eta_{ij}=\textrm{diag}(-1,+1)$, $L^{Da'}(d)$ is related to $G^{Da'}(d)$ by the two-dimensional Lorentz transformation
\begin{displaymath}
L^{Da'}_i=\Lambda_i{}^j G^{Da'}_j,\qquad\Lambda=
\left(
\begin{matrix}
\operatorname{ch}\varphi & \operatorname{sh}\varphi \\
\operatorname{sh}\varphi & \operatorname{ch}\varphi
\end{matrix}
\right)\in \mathrm{SO}(1,1),
\end{displaymath}
where $\varphi$ is the spectral parameter and $\ell_1=\operatorname{ch}\varphi$, $\ell_2=\operatorname{sh}\varphi$.

The zero-curvature representation of the Hamiltonian equations of tensile string seems have not been considered so far. Below, we give such representation and relate it to the Lax representation for the Hamiltonian equations of null string.

In the canonical approach, momentum density conjugate to the space-time coordinates is determined by the relation
\begin{gather}\label{16}
-\frac{1}{T\sqrt{-\gamma}}p_{m'}(\tau,\sigma)=\gamma^{\tau\tau}g_{m'n'}(X)\partial_\tau X^{n'}+\gamma^{\tau\sigma}g_{m'n'}(X)\partial_\sigma X^{n'}.
\end{gather}
Like in the tensionless case, there can be introduced tangent-space momentum density $p_{a'}(\tau,\sigma)=E_{a'}^{n'}p_{n'}$. In terms of it, one can express two first-class constraints that generate reparametriza\-tions of the tensile string world sheet
\begin{displaymath}
p_{a'}\eta^{a'b'}p_{b'}+T^2G^{Da'}_\sigma\eta_{a'b'}G^{Db'}_\sigma\approx0,\qquad p_{a'}G^{Da'}_\sigma\approx0.
\end{displaymath}
In the tension-to-zero limit, they turn into the constraints of the null string \eqref{10}. Lagrangian~\eqref{14} in terms of canonical variables has the form
\begin{gather}\label{17}
\mathscr L=p_{a'}G^{Da'}_\tau+\frac{\lambda}{2}\bigl(p^2+T^2G^{Da'}_\sigma\eta_{a'b'}G^{Db'}_\sigma\bigr)+\mu p_{a'}G^{Da'}_\sigma,
\end{gather}
where $\lambda^{-1}=T\sqrt{-\gamma}\gamma^{\tau\tau}$ and $\mu=\gamma^{\tau\sigma}(\gamma^{\tau\tau})^{-1}$ play the role of Lagrange multipliers at the first-class constraints.
These relations can be considered as two equations for the ratios of the components of inverse metric $\gamma^{ij}$. They can be solved to obtain appropriate for taking $T\to0$ limit parametrization of the Weyl-invariant density of inverse metric
\begin{displaymath}
 T\sqrt{-\gamma}\gamma^{ij}=\frac{1}{\lambda}\left(
\begin{matrix}
1 & \mu \\
\mu & \mu^2-T^2\lambda^2
\end{matrix}
\right)\;\xrightarrow[T\to0]{}\;
\frac{1}{\lambda}\left(
\begin{matrix}
1 & \mu \\
\mu & \mu^2
\end{matrix}
\right)=\rho^i\rho^j,
\end{displaymath}
where the world-sheet vector density introduced in \eqref{0} has the following expression via the Lagrange multipliers $\rho^i=\lambda^{-1/2}(1,\mu)$. Using it, one can express the Lagrange multipliers in terms of the vector density components
\begin{gather}\label{17'}
\lambda=\frac{1}{(\rho^\tau)^2},\qquad\mu=\frac{\rho^\sigma}{\rho^\tau}.
\end{gather}

From the tensile string action with the Lagrangian \eqref{17}, there follow two non-trivial equations
\begin{gather}
p^{a'}=-\frac{1}{\lambda}\bigl(G^{Da'}_\tau+\mu G^{Da'}_\sigma\bigr),\label{18}
\\
\partial_\tau p^{a'}+2G_\tau^{a'}{}_{b'}p^{b'}+\partial_\sigma\bigl(\mu p^{a'}+T^2\lambda G^{Da'}_\sigma\bigr)+2G_\sigma^{a'}{}_{b'}\bigl(\mu p^{b'}+T^2\lambda G^{Db'}_\sigma\bigr)=0.\label{19}
\end{gather}
Equation~\eqref{18} is another form of the definition of string momentum \eqref{16}. It is used to express components of the Lax connection $L(d)=d\tau L_\tau+d\sigma L_\sigma$ in terms of canonical variables
\begin{gather}
-L^{Da'}_\tau=\ell_1\bigl(\lambda p^{a'}+\mu G^{Da'}_\sigma\bigr)+\ell_2\left(\frac{\mu}{T}p^{a'}+T\lambda G^{Da'}_\sigma\right),\qquad L^{Da'}_\sigma=\ell_1 G^{Da'}_\sigma+\ell_2\frac{1}{T}p^{a'},\nonumber \\
L^{a'b'}_\tau=G^{a'b'}_\tau,\qquad L^{a'b'}_\sigma=G^{a'b'}_\sigma. \label{19'}
\end{gather}
The above components of the Lax connection enter the zero-curvature condition \eqref{15} that in component form reduces to single equation
\begin{gather}\label{20}
\partial_\tau L_\sigma-\partial_\sigma L_\tau+[L_\tau,L_\sigma]=0.
\end{gather}
It holds on dynamical equations \eqref{19} and the Maurer--Cartan equations \eqref{2} and \eqref{3} expressed in terms of the phase-space variables of the string
\begin{gather*}
\begin{split}
&\partial_\tau G^{Da'}_\sigma+\partial_\sigma\bigl(\lambda p^{a'}+\mu G^{Da'}_\sigma\bigr)+2G_\tau^{a'}{}_{b'}G^{Db'}_\sigma+2G_\sigma^{a'}{}_{b'}\bigl(\lambda p^{b'}+\mu G^{Db'}_\sigma\bigr)=0, \\
& \partial_\tau G^{a'b'}_\sigma-\partial_\sigma G^{a'b'}_\tau+2G_\tau^{a'}{}_{c'}G^{c'b'}_\sigma-2G_\sigma^{a'}{}_{c'}G^{c'b'}_\tau +2s\lambda\bigl(p^{a'}G^{Db'}_\sigma-p^{b'}G^{Da'}_\sigma\bigr)=0.
\end{split}
\end{gather*}

In the tension-to-zero limit, it is readily shown using \eqref{17'} that equation~\eqref{18} translates into equation~\eqref{11} defining null string momentum density and equation~\eqref{19} goes to dynamical equation of the null string \eqref{12}. Relation between the zero-curvature condition for the Lax connection \eqref{20} and the Lax equation for null string \eqref{7} is not straightforward. Equation~\eqref{7} holds solely on the null string equations and does not require utilization of the Maurer--Cartan equations. In the zero-curvature condition \eqref{20}, the part independent of the spectral parameter and that linear in $\ell_1$ are satisfied on the Maurer--Cartan equations, whereas the part linear in~$\ell_2$ is satisfied on dynamical equations of the string. Since in the $T\to0$ limit these equations reduce to dynamical equations of the null string, it is this part of the zero-curvature condition that reduces to the Lax equation \eqref{7}.
Assuming that the spectral parameter's dependence on~tension is such that \smash{$\varphi(T)\xrightarrow[T\to0]{}0$}, so that $\ell_1\to1$, $\ell_2\to0$ but $\ell_2/T$ remains fixed the limiting values of the pieces of the Lax connection \eqref{19'} proportional to $\ell_2$ are
\begin{displaymath}
-L^{Da'}_\tau\vert_{\ell_2}\;\xrightarrow[T\to0]{}\;\mu p^{a'},\qquad L^{Da'}_\sigma\vert_{\ell_2}\;\xrightarrow[T\to0]{}\; p^{a'}. \end{displaymath}
Using \eqref{17'} and combining them in the zero-curvature condition with the terms that contain the $\mathrm{SO}(1,D-1)$ generators, it can be shown that the part proportional to~$\ell_2$ indeed reduces to the Lax equation \eqref{7} with the Lax pair component $L$ expressed via the momentum of the null string~\eqref{12'}.

\subsection{Twistor formulations of (null) strings in anti-de Sitter space}

In conclusion, let us consider matrix realization of the coset element $\mathscr G$ that defines Cartan forms. This will make it possible to present discussed formulation for the null string in anti-de Sitter space, which isometry group is realized as conformal symmetry of the boundary Minkowski space, in terms of twistors \cite{Penrose1967,Penrose1972,Penrose1986}. (Formulation of the null strings in 4-dimensional Minkowski space in terms of $\mathrm{SU}(2,2)$ twistors was presented in \cite{Ilyenko2001a,Ilyenko2001b}.) Consider irreducible $d_R$-dimensional representation $R$ of the $\mathfrak{so}(2,D-1)=\mathfrak{ads}_D$ algebra. In case of $D=5$, four-dimensional fundamental representation corresponds to the known twistors of the anti-de Sitter space \cite{Adamo2016,Arvanitakis2016,Arvanitakis2017,Bars2004,Bars2005,Cederwall2000,Cederwall2004,Claus1999,Joung2024,Koning2024,Uvarov2018,Uvarov2019}. For other values of $D$, fundamental representations correspond to their generalizations \cite{Arvanitakis2016,Arvanitakis2017,Cederwall2000,Joung2024,Koning2023,Kuzenko2021}. Other finite-dimensional representations of the $\mathfrak{so}(2,D-1)$ algebra can be built via the tensor product of fundamental representation with itself and/or with the antifundamental one (if it is not isomorphic to the fundamental representation). Therefore, these representations can be viewed as corresponding to twistors of higher valences similarly to the case of $\mathrm{SU}(2,2)$ twistors \cite{Penrose1986}. (Such interpretation applies not only to the anti-de Sitter space but also to realization of the $(D-1)$-dimensional Minkowski space as the coset manifold of the (spinor covering of) conformal group on its subgroup generated by $\mathrm{SO}(1,D-2)$ rotations, dilatations and conformal boosts.) Then the coset element~$\mathscr G$ in \eqref{1} is given by $d_R\times d_R$ matrix $\mathscr G^\alpha{}_{(\beta)}$, $\alpha,(\beta)=1,\dots, d_R$, which upper index labels the twistor components and the index in brackets labels twistors themselves. Taking into account that $\det \mathscr G\not=0$, these $d_R$ twistors form complete basis in respective twistor space. The upper index is acted upon by the left global transformations from the (covering) of the~${\mathrm{SO}(2,D-1)}$ isometry group of the $D$-dimensional anti-de Sitter space. The lower index labels twistor basis components that transform under right local rotations from the (covering) of the stability group $\mathrm{SO}(1,D-1)$. It is assumed that the dimensions of respective representations of left and right groups equal $d_R$. Explicit dependence of $\mathscr G^\alpha{}_{(\beta)}$ on coordinates of the anti-de Sitter space determines incidence relations between components of the (higher-valence) twistors and these coordinates. The inverse of $\mathscr G^\alpha{}_{(\beta)}$ that enters definition of Cartan forms \eqref{1} is expressed via its transposed and/or complex conjugate. For instance, for 5-dimensional anti-de Sitter space with the isometry group $\mathrm{SO}(2,4)$, its double cover is isomorphic to the $\mathrm{SU}(2,2)$ group of unitary symmetry, for which $\mathscr G^{-1}=\bar{\mathscr G}=J\mathscr G^\dagger J$ with $J$ similar to the $\mathrm{diag}(+1,+1,-1,-1)$ matrix. This implies the constraint $J\bar{\mathscr G}J\mathscr G=I$ that should be taken into account, when the matrix $\mathscr G$ is considered as variational variable in the action.

In terms of the above considered twistor matrices, Lagrangian of the null string \eqref{4} is written~as
\begin{displaymath}
\mathscr L=-\frac12\rho^i\rho^j\operatorname{Tr}\bigl(\bigl(\bar{\mathscr G}\partial_i\mathscr G\bigr)\vert_{g/h}\bigl(\bar{\mathscr G}\partial_j\mathscr G)\vert_{g/h}\bigr),
\end{displaymath}
where $g=\mathfrak{so}(2,D-1)$ and $h=\mathfrak{so}(1,D-1)$, or in more explicit form
\begin{displaymath}
\mathscr L=-\frac12\rho^i\rho^j\operatorname{Tr}\bigl(P_{g/h}\bigl(\bar{\mathscr G}\partial_i\mathscr G\bigr)P_{g/h}\bigl(\bar{\mathscr G}\partial_j\mathscr G\bigr)\bigr).
\end{displaymath}
Here $P_{g/h}$ is the projector onto the quotient algebra $g/h=\mathfrak{so}(2,D-1)/\mathfrak{so}(1,D-1)$, expression for which depends on the representation $R$: $P_{g/h}R(M_{Dm'})=R(M_{Dm'})$, $P_{g/h}R(M_{m'n'})=0$. Taking into account that \smash{$\operatorname{Tr}(R(M_{Dm'})R(M_{Dn'}))=I^{-1}_R\eta_{m'n'}$}, one has
\[
G^{Dm'}(d)=I_R\operatorname{Tr}\bigl(R\bigl(M_D{}^{m'}\bigr)P_{g/h}\bigl(\bar{\mathscr G}d\mathscr G\bigr)\bigr),
\]
 where coefficient $I_R$ is the second Dynkin index of the representation. So, the null string Lagrangian can be written in another form
\begin{gather*}
\mathscr L=-\frac{I^2_R}{2}\rho^i\rho^j\operatorname{Tr}\bigl(R\bigl(M_D{}^{m'}\bigr)P_{g/h}\bigl(\bar{\mathscr G}\partial_i\mathscr G\bigr)\bigr)\eta_{m'n'}\operatorname{Tr}\bigl(R(M_D{}^{n'}\bigr)P_{g/h}\bigl(\bar{\mathscr G}\partial_j\mathscr G\bigr)\bigr).
\end{gather*}
It should be supplemented by the above discussed quadratic constraint for the matrix $\mathscr G$ with matrix Lagrange multiplier. This interpretation of the group-theoretic formulation of null strings in anti-de Sitter space-time in terms of twistors is readily extended to the case of non-zero tension
\begin{gather*}
\mathscr L=-\frac{TI^2_R}{2}\sqrt{-\gamma}\gamma^{ij}\operatorname{Tr}\bigl(R\bigl(M_D{}^{m'}\bigr)P_{g/h}\bigl(\bar{\mathscr G}\partial_i\mathscr G\bigr)\bigr)\eta_{m'n'}\operatorname{Tr}\bigl(R\bigl(M_D{}^{n'}\bigr)P_{g/h}\bigl(\bar{\mathscr G}\partial_j\mathscr G\bigr)\bigr).
\end{gather*}
It also applies to other models of point and extended objects with or without tension and also admits supersymmetric generalization.

\section{Conclusion}

Study of integrable structures allows us to unveil hidden symmetries that underlie dualities between gauge fields and strings in anti-de Sitter superspaces. In superspaces containing anti-de Sitter spaces of dimension $D<5$ it is obscured by broken supersymmetries. Study of simplifying limits such as those of zero and infinite tension of superstrings can provide important information on the integrable structures involved in respective dualities.

Here we obtained the Lax representation of equations of the null string in $D$-dimensional (anti-)de Sitter space and compared it with the zero-curvature representation of tensile string equations. For instance, for $D=4$ it has the same group-theoretic structure as the contribution proportional to generators of isometry algebra of the anti-de Sitter space to bosonic limit of the Lax representation of equations of massless superparticle on the $\mathrm{OSp}(4|6)/(\mathrm{SO}(1,3)\times \mathrm{U}(3))$ supercoset manifold and in the $\mathrm{AdS}_4\times\mathbb{CP}^3$ superspace. The obtained Lax representation can be readily generalized to null superstrings on supercoset manifolds. Generalization to $AdS$ superspaces containing supergroup manifolds, like the $\mathrm{AdS}_4\times\mathbb{CP}^3$ one, is non-trivial, but can give certain information on the integrable structure of tensile superstrings in these superspaces. This follows from the earlier found correspondence between the structure of the Lax pair of massless superparticle and the Lax connection of two-dimensional sigma-model on supergroup manifolds.

Another result of this work is the twistor interpretation of the formulation of (null) strings in anti-de Sitter space in terms of group-theoretic variables.

\subsection*{Acknowledgements}

It is a pleasure to thank A.A.~Zheltukhin for valuable discussions and careful reading of the manuscript.
Thanks also go to anonymous
referees for their remarks and suggestions that helped to improve the paper. Partial support by the STCU grant 9918 is gratefully acknowledged.

\pdfbookmark[1]{References}{ref}
\LastPageEnding

\end{document}